\documentclass[12pt,preprint]{aastex}

\shorttitle{Transparency Loss by Lyman_\alpha Clouds}
\shortauthors{ Schild and Dekker}

\begin{document}

\title{The Transparency of the Universe Limited by $Ly-\alpha$ Clouds }

\author{Rudolph E. Schild }
\affil{Center for Astrophysics,
        60 Garden Street, Cambridge, MA 02138}
\author{Marius Dekker}
\affil{2208 NE 153 Ave, Vancouver WA 98684}
\email{rschild@cfa.harvard.edu}

\begin{abstract} The brightnesses of supernovae are commonly
understood to indicate
that cosmological expansion is accelerating due to dark energy. 
However the entire discussion presumes a perfectly transparent universe
because no effects of reddening associated with the interstellar extinction
law are seen. We note that with two kinds of dark matter (baryonic and
non-baryonic) strongly dominating the known mass of the universe, it is
seriously premature to assume that these dark matter components have not
reduced the transmission of the universe for cosmological sources.

We show that the long-known $Lyman-\alpha$ clouds, if nucleated by the
population of baryonic dark matter primordial planetoids
indicated by quasar microlensing, would act as spherical lenses and
achromatically fade cosmologically distant sources. 
We attempt to estimate the amount of this cosmological fading, but
ultimately the calculation is limited by lack of a satisfactory model for
the tenuous outer parts of a primordial planetoid. We also consider
the effects of such cosmological fading on the light of quasars.

\end{abstract}

\keywords{ Galaxy: halo \--- baryonic dark matter: Dark Energy: Lyman-alpha
                   Clouds }

\section{Introduction}

Fundamental to our understanding of the nature of our universe from
observations of brightnesses of cosmologically distant objects is the
assumption that the universe is basically transparent, with any departures
from total transparency signaled by a reddening effect caused by scattering
of light by dust. However, with almost all of the matter in the universe in
2 unknown dark matter forms, there is the possibility that additional 
processes may be limiting the transparency. In the following, we propose
that achromatic refraction by the $Ly-\alpha$ clouds long known to pervade
intergalactic space may cause an effective reduction of the transparency.

The properties of the network of $Ly-\alpha$ clouds have been recently
reviewed by Rauch (1998). We remind here that the earliest
investigations of the clouds produced the conclusion that they must
dissipate on a short time scale by simple diffusion/evaporation, and that
therefore there must be some hot inter-cloud medium pervading all space to
confine them. Subsequent observations showed that the hot intercloud medium
does not exist, leaving the confinement of the clouds a mystery.

Recent developments suggest another possibility; the clouds do indeed
evaporate continuously but are continually refreshed by the planet-mass
primordial baryonic dark matter objects discovered in quasar microlensing
(Schild, 1996; 2004a,b). Such objects would produce a new perspective on the
transparency of the universe, because any centrally condensed spherical
object becomes a refracting lens, which would necessarily refract light
out of the beam of a cosmologically distant
background source. Because of the simple
molecular and atomic structure of hydrogen gas, the refraction would be
nearly achromatic, thus leaving no observable color signature.

In the remarks to follow, we offer two calculations of the refractive
properties expected for the cloud population. In section 2 we examine the
process of refraction applied to the $Ly-\alpha$ clouds if they are
nucleated by a compact object, and we check this calculation in section 3
with known properties of the terrestrial atmosphere. Here we notice that
dark objects of approximately this outer density and mass have been 
observed elsewhere
in the astronomical zoo, which gives us the size of the dark matter object
without any complicating model calculation.
Thus in section 4 we can easily determine
the cosmological loss of transmission due to this cloud population and in
section 5 we suggest falsifiable predictions implied by the mechanism. 
We conclude in section 6 that the light of quasars should also be affected.

\section{ The refracting layer density from the lens-maker's equation}

We adopt the simple picture of the $Ly-\alpha$ cloud (hereinafter Cloud)
as a spherically symmetrical object with an unknown density gradient and a
total mass of approximately $10^{-6} M_\odot$, after the quasar 
microlensing results
of Schild (1996). It will become apparent that the refraction occurs in
tenuously held outer layers of the object, which does not concern us.

The index of refraction of hydrogen at S.T.P. is 1.000132, and the
pressure and density dependence has been reported by Ruoff and Ghandehari
(1993) and by Stewart (1964) for the relevant temperature range 24 - 33 K.
The refractivity, (n-1), remains linear to very low densities measured.

The lensmakers equation can then be used to make an estimate of the focal
length of the lens, from which we determine the refracted angle. We use the
simplified version of the lensmaker's equation, also called the simple lens
equation, which is applicable because the refracted angle is extremely
small and the thickness of the Cloud, 2R, is small compared to the cosmological
distances involved. We treat the source as at infinite distance and compute
the focal length of the lens f as:

      $f = R/2(n-1)$

Here we have adopted the radius of the cloud as R. The most comparable
object in the astronomical zoo to our hypothetical Cloud is the knot found
abundantly in the Helix Nebula (Gibson, 2003). 
These objects have direct determinations
of mass, radius, and density from Meaburn et al (1998) and Huggins et al 
(1992) from direct CO detection, and
although they have been explained away as Rayleigh-Taylor instabilities in
the expanding gas shell, the measured density enhancement is much greater
than can be possible for such an instability. The fact that they probably
do not
show the expansion of the nebular shell (O'Dell and Handron, 1996), 
suggests that they
are our primordial baryonic dark matter objects.
For S.T.P, (Standard Temperature and Pressure)
the focal length of the Cloud computed for the $H_2$ index of
refraction is $2 x 10^{10} km$, and the refracted ray is at an angle of
approximately an arc minute..

For cosmological distances, refraction at a much
smaller angle suffices to remove light from the supernova or 
quasar image. A standard supernova or
luminous quasar accretion disc radius of $10^{15} cm$ (Schmidt and Wambsganss
2001) and the simple assumption that the refracting lens is half-way to the
cosmological source means that a refraction 
angle of approximately a nano-arcsec will displace the beam away 
from the observer.
This can be easily seen by imagining an observer at the quasar looking at
the Earth. If light beams are directed toward earth at angles with subtense
larger than the resolved quasar image, the light will not reach the observer.

A more careful calculation shows that the deflection needs to be more than
30 nano-arcsec for a luminous supernova shell of $R = 10^{15} cm$, or for a
quasar at the z=1.9 peak of the quasar redshift distribution.

Clearly the refraction angle needed to significantly fade the image of a
distant supernova or quasar is much smaller than that of our Cloud,
imagined above to be at S.T.P. If we take Cloud radii as above and a
refracting angle of 30 nano-arcsec and compute the density correction
implied by the refractivity required for our lens, we easily compute
a density of only $2 x 10^{-14} g cm^{-3}$ for the refracting layer in a
spherical Cloud lens of radius $2 x 10^{10} km$.

\section{The refracting layer density from scaled terrestrial refraction}

Another well-studied reference point for the refractive properties of a gas
cloud is the refraction of the setting Sun. It is a well known result
that the apparent horizon Sun is refracted 34 arcminutes from its
astronomical direction, for a terrestrial atmosphere at S.T.P. (Thomas and
Joseph, 1996). The theory of refraction in a nearly isothermal atmosphere
has been given by Lepetit and Lempel (2004) and Sloup (2003) and discussed
exhaustively by Garfinkel (1967). Thus it is
easy to scale to lower densities for the much smaller 30 nano-arcsec
deflection required here. Adopting an S.T.P.
 density of air of $.0013 g/cm^{3}$ and
noting that the index of refraction of hydrogen is half that of air
(Essen 1953)
we easily scale the density of a hydrogen atmosphere refracting by 30
nano-arcsec to have density $5 x 10^{-14} g/cm^3$. This is comfortably close to
the value surmised by a more standard Section 2 calculation.

We finally take the atmospheric density required to produce a refraction
effect large enough to remove the light from a supernova or quasar beam, 
$2 x 10^{-14} g/cm^{3}$, and ask what would be the radius of a Cloud to this
density value. In the absence of a model of a primordial hydrogen cloud
(brown dwarf) that has cooled for the entire life of the universe, we again
refer to the observations of such clouds in the Helix Nebula. Total masses
for the clouds, called cometary knots when ablated by the central star's
radiation pressure, have been investigated by direct CO mass measurement
(Huggins et al, 1992) and from an ablation theory applied to the
best resolved objects (Meaburn et al 1998). The estimated mass in both
methods is $10^{-5} M_\odot$, in near agreement with the value estimated from
quasar microlensing ($10^{-6} M_\odot$; Schild, 1996). 

The density computed above for our Cloud to have significant refraction 
is approximately $10^{-14} g/cm^{3}$. 
From Avagadro's number, this is equivalent
to $10^{10} molecules/cm^{3}$. This will be an important number to match to
the density profile of a primordial brown dwarf object to estimate the size
of such an object to this density. However we caution that at such low
densities, the atmosphere will be loosely bound to the planetoid and probably
irregular in structure.

We also note in caution that 
the measured diameter and total mass of the clouds have been combined 
in the cited references to
estimate the mean Helix knot density to be only $10^{6} particles/cm^{3}$, 
but these estimates will be dominated by the mass of the dense central
region (presumably having the density and pressure of the earth's core).
Thus, lacking a real model of such a primordial planetoid, our safest
recourse is to simply observe that the Helix Nebula knots are opaque to a
diameter of $2x10^{15} cm$ (Meaburn, 1996; O'Dell and Handron, 1996)
at all optical wavelengths. This is
the diameter we take for our calculation of the transparency of the
universe in the following section. Note that the size of a Cloud is
comparable to the size of a supernova or quasar optical disc $(10^{15}cm)$.

\section{The Transparency of the Universe limited by Baryonic Dark Matter
Objects} 

We now take the above estimates of the radius and mass of the dark matter
objects and compute their probable limits on the transparency of the
universe. We presume that any light incident upon the back side of an
object to the critical radius for refraction more than 30 nano-arcsec
is lost to the beam and eventually is absorbed or contributes to the
cosmological backgroud light. We assume that along the line of sight to a
cosmologically distant object at redshift 2, a Milky Way type galaxy
with a mass including its Baryonic dark matter halo of $10^{12} M_\odot$
(Allen's Astrophysical Quantities, 4'th edition, p571) and a size of 50 kpc
(distance to the Magellanic clouds) will be encountered. For our toy 
calculation, we assume that the refracting
objects are uniformly distributed in a cube $10^{5} pc$ on a side, and the
mass of each planetoid is $10^{-6} M_\odot$. Then the surface area 
covered by the
planetoid objects will be $10^{49} cm^{2}$ and the projected total area of the
Galaxy Halo will be $10^{50} cm^{2}$. Thus the refracting halos 
cover 10 percent
of the light path through the Galaxy Halo. and several galaxy halos along the
line of sight would refract a significant fraction of the light of a
cosmological source.

Thus according to this simple optical refraction model, 
the baryonic dark matter objects
discovered in quasar microlensing reduce the brightnesses of distant
cosmological sources.

Of course the simple picture needs refinement, but little improvement will
be possible until a primordial brown-dwarf type object with 
$10^{-6} M_\odot$ 
is modeled to see what its effective diameter is for refractive attenuation
of cosmologically distant background sources. For the present we only claim
that there is the possibility that refractive opacity to cosmologically
distant sources must be considered before concluding that supernova
brightness curves strongly indicate a second inflation event (Dark Energy)
in the universe.

\section{Some Falsifiable Predictions}

A universal population of refracting clouds has already been implied by
$Ly-\alpha$ absorption clouds long seen in quasar spectra. The present
interpretation of their refractive properties has two immediate
implications for observable side effects:

1. cosmologically distant sources should be surrounnded by faint halos
   caused by refracted light

2. since some fraction of the refracted light is not absorbed, it should
   contribute to a cosmological background light.

Both effects have been seen.

With respect to item 1, it might be argued that in a linear process about
as much light gets scattered into the quasar/supernova image as is
scattered away. However, observations of occultations of stars by solar
system planets show peaks in their brightness curves that indicate that
even for tightly held planetary atmospheres in the Solar System, 
layered structures exist
that make the refraction process complex and non-linear (Hubbard et al,
1972; Veverka et al, 1974). 

At the same time, halos are ordinarily seen around quasars, where the
underlying fuzz is interpreted to be the resolved image of the host galaxy,
also present. The study of the faint fuzz is complicated by the bright
quasar emission and the usual instrumental problems due to light scattering
in the spectrograph and telescope optics.
However the fiber optic bundle field resolving
spectrograph applied by Mediavilla et al (1998, and further
related references) shows a halo in AGN MKN 509 out to a distance of 8 kpc
from the galaxy center in the light of the broad emission lines originating
in the nucleus. The authors concluded that the broad emission line cloud
could not be
16 kpc in diameter and attributed the observation to light scattering, but
a process of refraction was not considered. Thus such small halos around
distant stellar sources are observed, and whether caused by refraction or
scattering they will reduce the measured brightness of cosmologically
distant sources.

Halos around supernova are unsurprising because supernovae are often
associated with clustered stars, particularly the Type 2 variants.
The profile fitting used to measure the supernova brightness is insensitive
to the existence of faint fuzz in the low S/N region surrounding the bright
stellar component.

The issue of cosmological background light has been controversial because
while its detection has been claimed, it is just a faint residual seen
against a bright foreground caused by the zodiacal light. Nevertheless in a
recent report, Dwek et al (2005) conclude that the
detected radiation is too bright to be understood by a hypothetical Pop III
sky background. Our refraction model predicts that the Extragalactic
Background Light should be comparable to, or slightly less than, the average
background light of all resolved sources.

A weak prediction comes from calculation of
the neutral hydrogen optical path length through the lensing
Cloud. The above numbers for the density and size of our cloud suggest
that the optical path through the cloud would produce a surface mass
density of approximately $10^{22} particles/cm^{2}$, which is the surface
mass density of the so-called damped Lyman-alpha systems. It would be
interesting to see whether cosmological sources with such damped
Lyman-alpha profiles in their ultraviolet spectra are slightly fainter
(about 0.1 magnitudes per damped profile) than otherwise.

\section{Transmission of the Universe to the Light of Quasars and
Supernovae} 

The study of quasars has long ago produced a puzzling observation that is
probably relevant. The number of quasars peaks
at a redshift around z = 1.9. This means that if we consider all directions
of space and ask how many quasars are contained as a function of redshift,
that function will peak at z = 1.9

There is presently no accepted theory of the formation of quasars, and this
peak in the number distribution is unexplained. However it can be easily
understood as resulting from the reduced transparency of the universe also
seen in the supernovae, and the simplest assumption that quasar intrinsic
luminosities have not changed to z = 5.
Recall that the supernova brightness - redshift relation purported to
demonstrate dark energy shows about a 1.4 magnitude deficit for the
quasar peak at z = 1.9. In other words, the dark matter cosmology
curve is fainter than the standard cosmology curve at the quasar density
peak by 1.4 magnitudes. This means that the transmission losses by the dark
matter are becoming significant at these redshifts, and the proposal is that
both the supernova brightness deficit and the quasar brightness peak can
reasonably be understood as resulting from reduced transmission of the
universe due to the baryonic dark matter.

Moreover, the exact form of the reduced transmission law exactly fits the
supernova observations. It has long been known (Zuo and Lu, 1993)
that the density of $Ly-\alpha$ clouds increases with redshift as 
$(1+z)^{2.8}$. It has also been noticed by Goobar et al (2002) that
an absorption law scaling as $(1+z)^{3}$ and constant beyond z = .5, 
can explain
the supernova brightness-redshift relationship of Riess et al (2004). 
Thus it is easy to understand how the known $Ly-\alpha$ 
clouds can exactly describe the supernova brightness anomaly presently
ascribed to dark energy.

\end{document}